# On Negative Index Metamaterial Spacers and Their Unusual Optical Properties

**Muhammad I. Aslam,**[1,2] **Durdu Ö. Güney,**[1,*]

[1]*Department of Electrical and Computer Engineering, Michigan Technological University, Houghton MI 49931, USA*
[2]*Department of Electronic Engineering, NED University of Engineering and Technology, Karachi 75270, Pakistan*
*\*Corresponding author: dguney@mtu.edu*

We theoretically investigate the possibility of using a metamaterial structure as a spacer, named as metaspacer, which can be integrated with other materials in microfabrication. We show that such metaspacers can provide new optical behaviors that are not possible through conventional spacers. In particular, we investigate negative index metaspacers embedded in fishnet metamaterial structures and compare them with conventional fishnet metamaterial structures. We show that the negative index metaspacer based fishnet structure exhibits intriguing inverted optical response. We also observe that the dependence of the resonance frequency on the geometric parameters is reversed. We conclude with practicality of these metaspacers.

## 1. INTRODUCTION

Following the studies on backward wave propagation [1-4], in 1968 Veselago systematically showed that media with simultaneously negative permittivity and permeability, referred later as negative index metamaterials, exhibit unusual and interesting properties such as negative refraction, reverse Doppler's effect, and reverse Cherenkov effect [5]. Research on metamaterials did not gain much attention until Pendry, et al. proposed structures to artificially achieve negative permittivity [6,7] and negative permeability [8]. This was followed by actual fabrication of negative index metamaterials by Smith et al [9,10]. Since then, numerous metamaterials have been proposed for different frequency regimes [11-14] to demonstrate many interesting properties and applications such as high precision lithography [15], perfect lens [16], high resolution imaging [16,17], invisibility cloaks [18], small antennas [19], optical analog simulators [20,21], and quantum levitation [22].

On the other hand, conventional materials that are used as spacers in microfabrication provide inherently limited optical and electronic properties. For example, dielectric spacers have permittivity higher than unity and are generally nonmagnetic. However, metamaterials can be designed to provide almost any value of permittivity ($\varepsilon = \varepsilon' + j\varepsilon''$) and permeability ($\mu = \mu' + j\mu''$). Utilizing this feature, metamaterials can replace conventional spacers in microfabrication. Such 'metaspacers' may be used in applications requiring very low index materials [23] or high permeability ferrites [24]. Furthermore, metaspacers can lead to devices/applications requiring spacers having index less than unity (even negative). Despite this great potential in microfabrication of new electromagnetic devices, metaspacers have not been studied. Here we define a metaspacer as a metamaterial structure that can be used as a spacer, integrated with other materials to fabricate novel devices that are not possible by using conventional materials.

Metaspacers can be designed to support different types of surface plasmon polariton (SPP) modes. A naturally available dielectric and a metal can form an interface with opposite signs of permittivity and support p-polarized SPPs. However, using a metaspacer, an interface with different signs of permittivity and/or permeability can be realized to support p- (or TM-) and/or s- (or TE-) polarized SPPs, respectively [25, 26]. For example, an interface between a metal and a negative index material can support s-polarized SPPs.

## 2. IDEALIZED METASPACER

In this letter, we theoretically investigate metaspacers as replacement to conventional materials to extend the capabilities of the devices produced by microfabrication. Our letter seeks answer to the question, "Can we make new materials (i.e., meta-metamaterials) with novel physical properties using metamaterials made out of natural materials?" In particular, we choose to study negative index metaspacer (most unusual spacer) embedded fishnet metamaterial structure (most studied and convenient optical metamaterial structure). The negative index metaspacer is placed between two metal layers. We compare the results with the (dielectric-based) conventional fishnet metamaterial structure (referred as CFS for short, "conventional" in the sense that the structure incorporates a non-dispersive spacer with a positive refractive index). We used frequency domain analysis of commercially available COMSOL Multiphysics for all simulations. Single unit cell of the fishnet structure used in the analysis is shown in Fig. 1. A square lattice having a period $p = 400$nm with square holes of side length 250nm is used. We considered one functional-layer of the fishnet structure consisting of three layers (silver-spacer-silver). The thickness of each metal layer is assumed to be $t = 35$nm and that of the spacer is $s = 30$nm. Silver is modeled by the Drude model with the bulk plasma frequency $f_p$ =2180THz and the collision frequency $f_c$ =13.5THz [27] (the reader is referred to a recent detailed quest [28] for low-loss passive optical materials for metamaterials and plasmonics). Due to the symmetry, the optical response is independent of incident parallel polarizations. We used the field configuration shown in Fig. 1, where **E**, **H**, and **k** denote electric field, magnetic field, and wave vector, respectively. In the negative index metaspacer embedded fishnet metamaterial structure (i.e., unconventional fishnet structure, referred as UFS for short) simulations, the metaspacer is initially modeled as an idealized lossless and non-dispersive negative index material followed by a discussion on practical realization.

We observed that a metaspacer with a high absolute value of negative permeability is required to achieve negative index using UFS. Therefore, the metaspacer is modeled by $\varepsilon = -2.5$ and $\mu = -5.0$. In order to have a good comparison with the CFS, the dielectric in the CFS is modeled by $\varepsilon = 2.5$ and $\mu = 5.0$. The background material is assumed to be air ($\varepsilon = \mu = 1.0$). Effective material parameters [$\varepsilon$, $\mu$, and index ($n = n' + jn''$)] are retrieved, from simulated reflection and transmission coefficients, using the isotropic retrieval procedure [29]. The retrieved effective parameters as well as the transmittance (T), reflectance (R), and absorption (A=1–T–R) are plotted in Fig. 2, for both fishnet structures. It can be observed that the maximum transmittance of the CFS in the negative index band is 0.44, in contrast to 0.69 achieved through the UFS. Similarly, for the latter, the reflectance drops down to 0.02 in contrast to 0.11 achieved in the former.

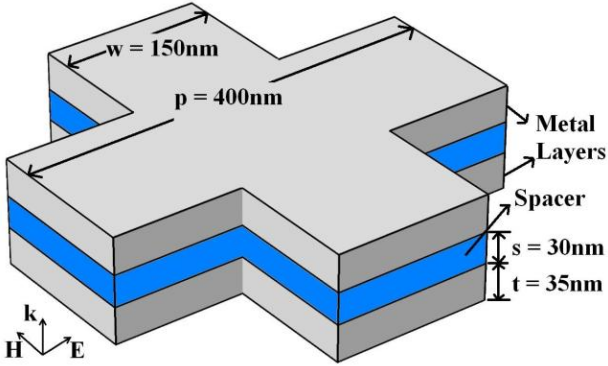

Fig. 1. (Color online) Unit cell of the simulated fishnet metamaterial structure.

The retrieved index for the CFS exhibits a negative index band extending from 263THz to 329THz with a minimum value of –3.13 having a full width at half maximum (FWHM) of 15THz. The figure of merit (FOM = $-n'/n''$) at the operating frequency (304THz) is found to be 2.95. On the other hand, the negative index band for the UFS extends from 240THz to 363THz with a minimum value of –3.0 and a FWHM of about 42THz. The FOM is 7.15 at the operating frequency (294THz). Operating frequency in this letter refers to the frequency corresponding to $n' = -1$. Although the operating frequencies of both structures are very close, the FOM and the negative index bandwidth of the UFS are about 2.5 times higher. The magnetic resonance in the UFS is also stronger and wider. The $\mu'$ reaches a minimum value of –3.85 whereas the minimum for the CFS is only –1.48. Fig. 3 shows the $y$-component of the magnetic field [Figs. 3(a) and 3(c)] and the $x$-component of the current density distributions [Figs. 3(b) and 3(d)] at the magnetic resonance frequencies of 297THz and 324THz for the CFS and the UFS, respectively. The other components were not shown for clarity since they are not relevant for the observed magnetic resonance. The magnetic fields are generated mainly inside the spacers and the currents are induced in the metallic layers as expected. The magnetic field in the central cross-sections of the UFS unit cell is significantly stronger than that of the CFS unit cell, although the fields are comparable at the outer cross-sections [i.e., compare Fig. 3(a) with Fig. 3(c)]. Furthermore, there is no important contribution to the magnetic response from the central cross-sections of the CFS unit cell, since the magnetic field distribution in these regions is approximately an odd function [see Fig. 3(a)]. However, there is important contribution to the magnetic response from the central cross-sections of the UFS unit cell [see Fig. 3(c)]. These observations can also be confirmed by comparing the current density distributions in the CFS and the UFS [i.e., compare Fig. 3(b) with Fig. 3(d)]. Therefore, consistent with the retrieved effective parameters in Fig. 2(a), the UFS exhibits stronger magnetic response than the CFS.

Furthermore, the double-negative pass-band region (i.e., the region where $\varepsilon'$ and $\mu'$ are simultaneously negative) for the UFS exists at the low-frequency side of the negative index band as opposed to high-frequency side for the CFS. Similar "inverted" behavior is also observable in the retrieved effective permeability. The "shapes" of the permeability curves are qualitatively similar for both cases except that the curves for the UFS are flipped around the resonance frequency. This inverted optical response results in two interesting consequences: (i) permeability similar to ferrites at microwave frequencies [30], (ii) simultaneously negative group and phase velocity at the low-loss region. The latter was observed recently in an experiment in the high-loss region (i.e., single negative region) [31, 32].

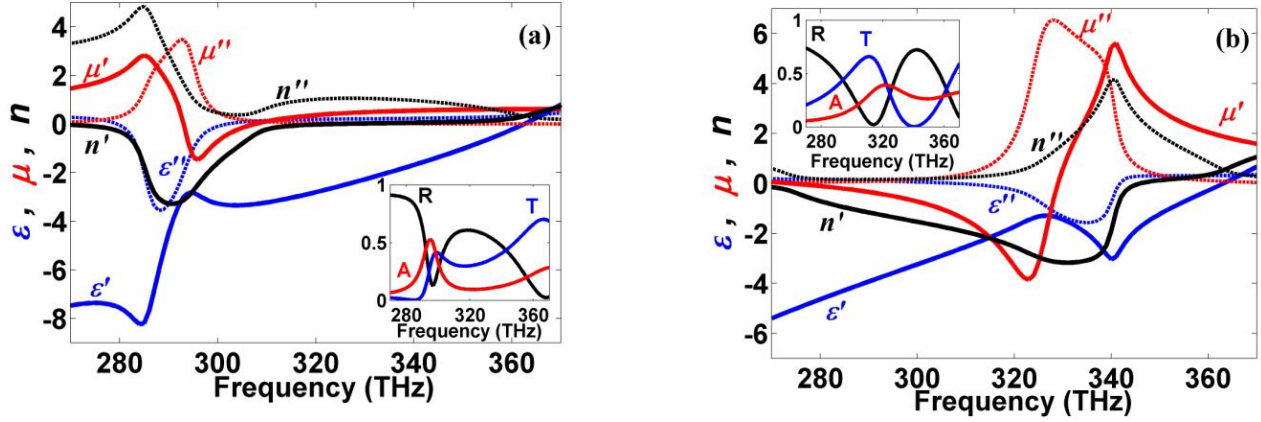

Fig. 2. (Color online) Retrieved effective parameters for the (a) CFS and the (b) UFS. Insets show the corresponding T-R spectra.

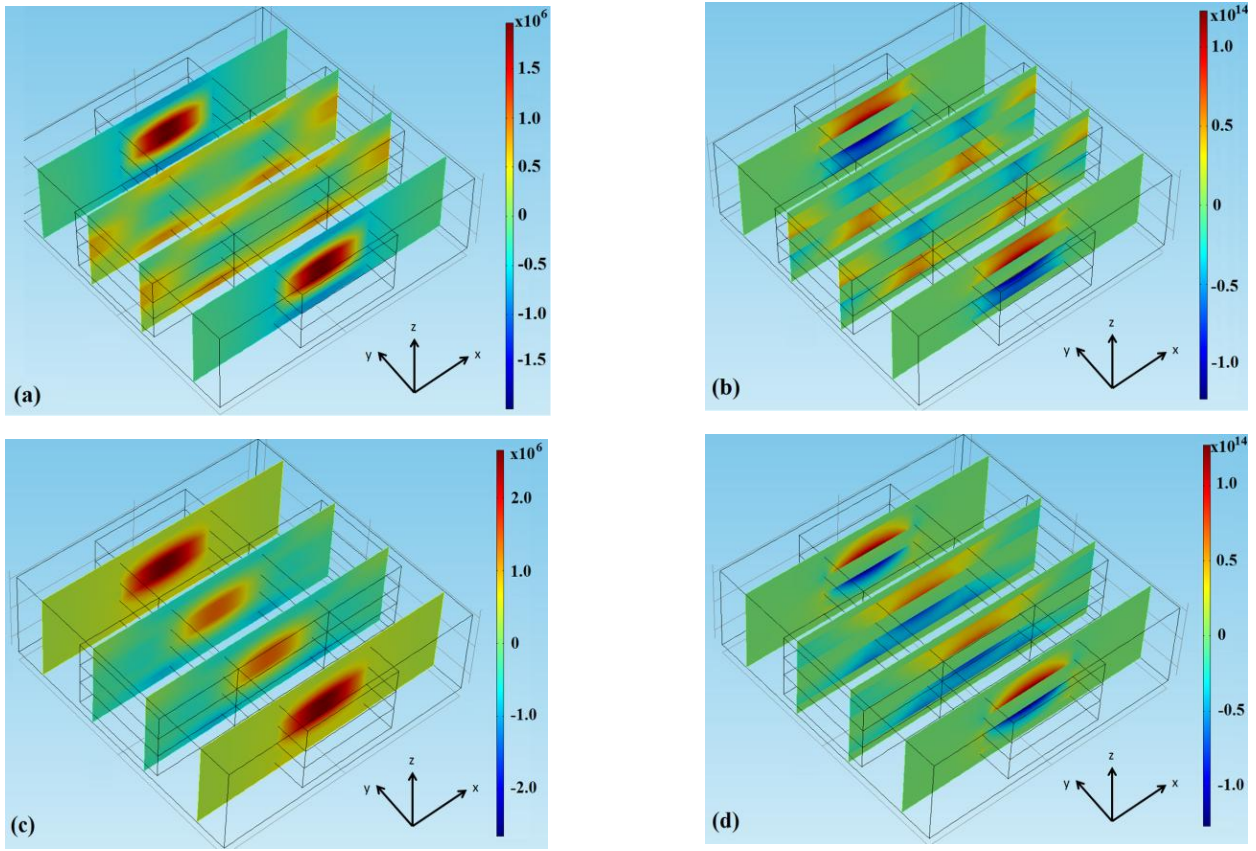

Fig. 3. (Color online) (a) Plot of the $y$-component of the magnetic field (i.e., parallel to the incident magnetic field) at four different cross-sections of the CFS along the $y$-axis, and (b) $y$-component of the corresponding current distribution. (c) Respective magnetic field and the (d) current distribution for the UFS. The plots correspond to the magnetic resonance frequencies of 297THz for the CFS and 324 THz for the UFS. Colors show the magnitude and direction of the magnetic field or current density [i.e., red corresponds to $+y$ ($+x$) and blue corresponds to $-y$ ($-x$) direction for the magnetic field (current density)].

We also notice that the dependence of the operating frequency on the physical geometry is reversed as well for the UFS. For example, while the operating frequency of the CFS increases with the number of functional layers [33], the operating frequency of the UFS reduces with the increasing number of functional layers. For instance, operating frequencies for the two- and three-functional-layer-UFSs for the geometry shown in Fig. 1, are 259THz and 243THz, respectively. Additionally, we observed that increasing the spacer thickness reduces the resonance frequency of the UFS whereas this increases the resonance frequency in the CFS. The latter is especially important for the homogeneous effective medium (HEM) approximation requirement of metamaterials because the required theoretical spacer thicknesses at optical frequencies can easily be comparable to free-space wavelength, hence violating the HEM approximation.

The CFS exhibits negative index when the magnetic resonance, due to the SPP excitation, occurs at the frequency close to the diluted plasma frequency [34]. The alternating layers of this structure form interfaces with opposite signs of permittivity [metal ($\varepsilon' < 0$) and dielectrics ($\varepsilon' > 0$)]; hence they support p-polarized SPPs [25, 26, 34, 35]. Therefore, the negative permeability in the CFS arises from mainly the p-polarized SPPs. In contrast, alternating layers of the UFS provide opposite signs of permeability. The metal is characterized by the negative permittivity and positive permeability ($\mu = 1.0$) while the negative index metaspacer is characterized by simultaneously negative permittivity and permeability. Therefore, at the interface, only the permeability has opposite signs. These types of interfaces support s-polarized SPPs [25, 26]. Therefore, the magnetic response in the UFS arises from mainly the s-polarized SPPs. It must be noted that the UFS also supports p-polarized SPPs due to the interfaces between air and metal/negative index metaspacer. However, because the magnetic response is generated by the metal-metaspacer interfaces parallel to the incident magnetic field, the UFS can be regarded as being driven by the s-polarized SPPs.

## 3. INFLUENCE OF DISPERSION AND LOSSES ON METASPACERS

The s-polarized SPPs have different dispersion relation and properties than the p-polarized SPPs [25]. The attenuation coefficients associated with the s-polarized SPPs are usually smaller compared to those of p-polarized SPPs [26]. This is also in accordance with our observation of higher transmission and higher FOM in the UFS. However, it must be noted that above simulations assume a loss-less and non-dispersive metaspacer. In practice, the negative index materials are inherently lossy and highly dispersive especially around the resonance frequency. Below we will study how losses and frequency dispersion influence our results shown in Fig. 2. However, we will not consider spatial dispersion and anisotropy, since the structure in question is designed strictly for normal incidence [36-40].

Practically low-loss negative index region for the metaspacer corresponds to simultaneously negative $\varepsilon$ and $\mu$ which are usually modeled as [25, 26]

$$\varepsilon_{MS} = 1 - \omega_p^2 / (\omega^2 + j\omega\gamma_e) \quad (1)$$

$$\mu_{MS} = 1 - F\omega^2 / (\omega^2 - \omega_0^2 + j\omega\gamma_m) \quad (2)$$

where $\omega_p$ is the bulk plasma frequency, $\omega_0$ is the magnetic resonance frequency, $\gamma_e$ and $\gamma_m$ are electric and magnetic damping coefficients, respectively, and $F$ is the filling ratio. These effective constitutive parameters are illustrated in Fig. 4(a) using arbitrary values in (1) and (2). The plot can be divided into four regions. For the regions ① and ③, the metaspacer has $\varepsilon' < 0$ and $\mu' > 0$, which are the same as that of the metal. Thus, no SPPs are excited at the corresponding frequencies. The metaspacer exhibits a flatter dispersion in region ④, and behaves like a dielectric. The optical behavior of the UFS in region ④ closely approximates the CFS with a non-dispersive spacer except that the metaspacer can be used to replace a dielectric with $\varepsilon$ and $\mu$ less than unity. The metaspacer exhibits double-negative pass band in a narrow range of frequencies corresponding to region ②. However, the metaspacer has highly dispersive permeability and is lossy in this region which makes the use of negative index metaspacer difficult.

In Fig. 4(b), we show the retrieved index for an UFS where the metaspacer parameters are set to exhibit the wideband double-negative response displayed in the inset (i.e., similar to region ②), by taking $\omega_p$=600THz, $\omega_0$=250THz, $\gamma_e$=1THz, $\gamma_m$=75THz, and $F$=3.25. These arbitrary values do not change the main conclusions drawn below. The imaginary part is ignored to illustrate only the impact of the dispersion. Two negative index bands are apparent in the retrieved results around 300THz. The low-frequency band on the left shows an index profile similar to that of the UFS with a non-dispersive negative index metaspacer except that it has narrower bandwidth.

In Fig. 5, we plot the retrieved effective parameters for a non-dispersive but lossy metaspacer. The metaspacer is modeled by $\varepsilon = -2.5 + j0.05$ and $\mu = -5 + j0.1$. This corresponds to a FOM of about 50 (i.e., moderately lossy metaspacer). The real parts of the constitutive parameters are the same as in Fig. 2(b). We observe that although the interesting features arising from the inclusion of the double-negative metaspacer still persist, the FOM for the UFS is reduced below that of CFS. How the imaginary parts influence the overall performance of the UFS, in comparison with the lossless UFS and CFS, are summarized in Table 1. We note that the lossy UFS has the widest double-negative bandwidth. We also found that negative index of the UFS completely vanishes and turn into a positive index (not shown) when the FOM of the metaspacer is further decreased down to 12 by increasing the imaginary parts of $\varepsilon$ and $\mu$. Indeed, larger values of FOM have been predicted in fishnet structures [33, 41]. We should also note that a recent paper by R. Paniagua-Dominguez, *et al* [42] shows a FOM of about 300 for an isotropic two-dimensional negative index metamaterial structure designed for near-IR wavelengths.

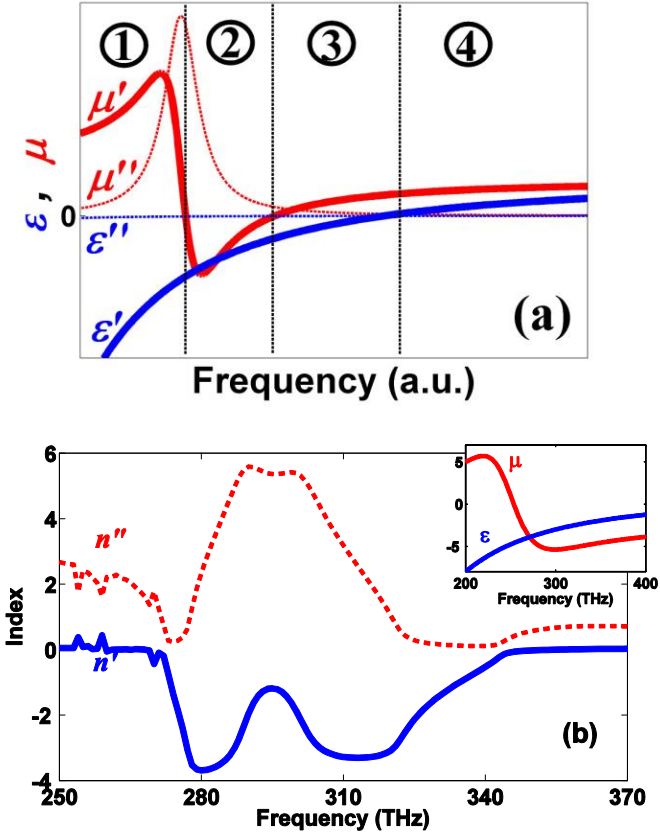

Fig. 4. (Color online) (a) An arbitrary plot for $\varepsilon$ and $\mu$ of metaspacer. (b) Retrieved index for the UFS where the metaspacer exhibits the response shown in the inset. Please see the text for the parameters used.

4. DISCUSSION

Above results suggest that, practical metaspacers (whether positive or negative index), ideally, require (i) sufficiently low-loss operation. Additionally, for negative index metaspacers (ii) wide negative index band, (iii) strong magnetic response, and (iv) relatively flat dispersion in permeability around the operating frequency are desired so that they could be approximated as non-dispersive materials with (unusual) properties superior to natural or conventional materials that are used in microfabrication. These properties can be realized at optical frequencies by the demonstration of broadband active metamaterials [43-46] and/or passive ultra low-loss negative index metamaterials [42], and at microwave frequencies by passive (i.e., using strong coupling and periodicity effects) or active [47] split-ring-resonator based metamaterials.

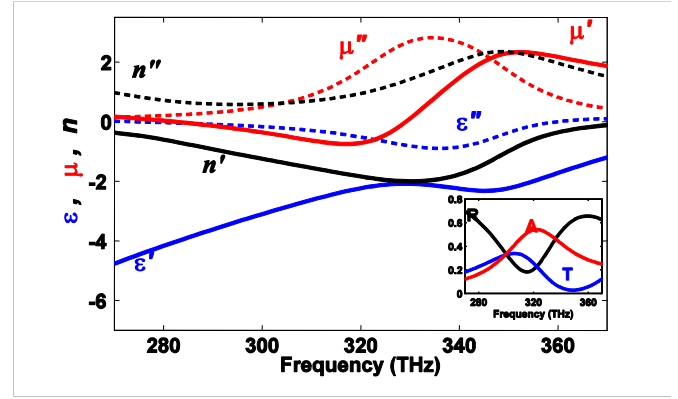

Fig. 5. (Color online) Retrieved effective parameters for non-dispersive lossy metaspacer. The metaspacer is modeled by $\varepsilon = -2.5 + j0.05$ and $\mu = -5 + j0.1$. The inset shows the T-R spectra.

How the interfaces will behave when actual discrete metamaterial structures are used as metaspacers will be studied in a future work. As long as the feature sizes of the metaspacer is sufficiently subwavelength along all relevant directions with respect to the operating free space and/or SPP wavelength, unwanted local fields at the interfaces can be eliminated. This imposes stringent conditions on the practical realization of metaspacers with effective medium response especially in optical frequencies. Therefore, nano-chemistry based bottom-up self-assembly approaches [48-55] may be inevitable for optical frequencies. However, without much trouble, the metaspacer concept that we introduced here can be also applied to low frequencies due to the scalability of electromagnetic waves. At low frequencies, there exist, for example, extremely subwavelength metamaterials, which can have unit cell sizes of about 2000 times smaller than the resonant wavelength [56]. These extremely subwavelength metamaterials can function as metaspacers by embedding in split-ring-resonator based metamaterials with relatively low unit cell size to free space wavelength ratio. This approach may be scaled up to low THz operating frequencies with current technology.

| Observation | CFS (Lossless dielectric) ε = 2.5, μ = 5 | UFS (Lossless Metaspacer) ε = –2.5, μ = –5 | UFS (Lossy Metaspacer) ε = –2.5 + j 0.05, μ = –5 + j 0.1 |
|---|---|---|---|
| Max (Transmittance) | 0.44 | 0.69 | 0.34 |
| Min (Reflectance) | 0.11 | 0.02 | 0.18 |
| Operating frequency | 304THz | 294THz | 293THz |
| FOM | 2.95 | 7.15 | 1.75 |
| Negative index band | 263-329THz | 263-329THz | 220-379THz |
| FWHM | 15THz | 42THz | 58THz |
| Min(n′) | –3.13 | –3.0 | –2.0 |
| Min(μ′) | –1.54 | –3.85 | –0.75 |
| Frequency at min(μ′) | 297THz | 326THz | 317THz |

Table 1. Summary of how the imaginary parts in ε and μ influence the overall performance of the structure. CFS, lossless UFS, and lossy UFS are compared.

5. CONCLUSION

In conclusion, unlike conventional materials [such as spacers, substrates, (non)engineered materials] commonly used in microfabrication, the properties of metaspacers can be almost arbitrarily controlled and optimized to manifest the desired optical properties to extend the functionality of available materials and devices. We introduced metaspacers for the first time in this letter. Particularly, we studied a negative index metaspacer embedded fishnet metamaterial structure which can support s-polarized SPPs. We have shown that this structure exhibits many intriguing features compared to CFS such as inverted optical response. We have also found that negative-index metaspacer reverses how the resonance frequency of the metamaterial structure depends on the geometric parameters such as spacer thickness and number of functional layers. Additionally, we studied how the losses and dispersion influence these features. Not surprising, we have found that the dispersive nature of the metaspacer reduces the bandwidth (over which the metaspacer exhibits interesting optical response such as "inversion") while the losses reduce the figure of merit. Therefore, loss and dispersion compensation to some extent may be necessary (by using, for example, active metamaterials and/or ultra low-loss metamaterials [42]), especially around the resonance frequency of the metaspacer. However, at sufficiently high frequencies, the adverse effects of the losses and dispersion can be neglected and thus passive metaspacers can be used as low-dielectric-constant spacers.